\documentclass[%
 aip,
 amsmath,amssymb,
 reprint,
]{revtex4-1}

\usepackage{graphicx}
\usepackage{dcolumn}
\usepackage{bm}
\usepackage{soul}
\usepackage{color}
\usepackage[utf8]{inputenc}
\usepackage[T1]{fontenc}
\usepackage{mathptmx}
\usepackage{etoolbox}

\makeatletter
\def\@email#1#2{%
 \endgroup
 \patchcmd{\titleblock@produce}
  {\frontmatter@RRAPformat}
  {\frontmatter@RRAPformat{\produce@RRAP{*#1\href{mailto:#2}{#2}}}\frontmatter@RRAPformat}
  {}{}
}%
\makeatother
\begin{document}

\preprint{AIP/123-QED}

\title{Combined spatially and temporally multiplexed photonic reservoir computer with a diffractively coupled VCSEL-array}
\author{J. Robertson}
 \affiliation{Institute of Photonics, SUPA, Dept. of Physics, University of Strathclyde, 16 Richmond St, G1 1XQ, Glasgow, UK
}
 
\author{M. Pfl\"uger}%
 \affiliation{Instituto de F\'{\i}sica Interdisciplinar y Sistemas Complejos, IFISC (UIB-CSIC), Campus Universitat de les Illes Balears, Ctra. de Valldemossa km. 7.5, 07122 Palma, Spain
}%

\author{I. Fischer}%
 \affiliation{Instituto de F\'{\i}sica Interdisciplinar y Sistemas Complejos, IFISC (UIB-CSIC), Campus Universitat de les Illes Balears, Ctra. de Valldemossa km. 7.5, 07122 Palma, Spain
}%

\author{M. C. Soriano}%
 \affiliation{Instituto de F\'{\i}sica Interdisciplinar y Sistemas Complejos, IFISC (UIB-CSIC), Campus Universitat de les Illes Balears, Ctra. de Valldemossa km. 7.5, 07122 Palma, Spain
}%

\author{A. Hurtado}
 \affiliation{Institute of Photonics, SUPA, Dept. of Physics, University of Strathclyde, 16 Richmond St, G1 1XQ, Glasgow, UK
}

\email{joshua.robertson@strath.ac.uk}

\date{\today}

\begin{abstract}
We report and analyse the classification performance of an experimental hybrid spatio-temporal photonic reservoir computer based upon a free-space VCSEL array.
We demonstrate experimentally the enhancement of spatial-only reservoir operation, featuring the diffractive coupling of lasers in an external cavity, by exploiting up to 88 virtual nodes with time multiplexing.
We analyse the dependance of performance on the spatial and virtual node number, and achieve an improvement for both spatial- and temporal-only reservoirs with a reduced test error of 0.026 in a classification task. 
Further, given the high bandwidth of the non-linear laser transformation, we demonstrate the expansion of a 12 spatial node network to a 968 node network, operating at an input time of 17.6\,ns, maintaining high processing speed and improving network scalability and performance.   
\end{abstract}

\maketitle

\section{\label{sec:level1}INTRODUCTION}

Reservoir computing (RC), an unconventional-computing paradigm for neural network implementation, is a promising route to alleviating the challenging energy efficiency and speed bottlenecks faced by Von Neumann computing architectures in the booming artificial intelligence (AI) era~\cite{yan2024emerging,nakajima2021reservoir}.
Reservoir computers are used to create simplified but powerful recurrent neural networks that exploit a rich feature space, expanding information to a high dimensional representation using various non-linear transformations~\cite{maass2002real,jaeger2004harnessing}.
In RC, the network manifests as a `reservoir' of nodes with fixed unknown connections and weights.
This feature simplifies the learning (training) requirements to a small set of output node weights (the readout layer), improving the efficiency and speed of system training~\cite{lukovsevivcius2009reservoir}.
Catalysed by innovations making learning concepts hardware-agnostic, neuromorphic computing has been thriving.
A core appeal of RC is therefore its deployability to physical hardware~\cite{tanaka2019recent}, and given the demand for energy efficient, high-speed computing, the photonic platform is being actively researched~\cite{van2017advances,abreu2024photonics}. 
This is giving rise to the field of Photonic Reservoir Computing (PRC)~\cite{Abdalla2025}.

Large bandwidth, low-loss passive components (e.g. waveguides and couplers), high-speed low-latency transmission and access to optical nonlinear elements is allowing for the photonic platform to create promising all-optical RC systems void of costly optical-electrical conversions.
The first demonstrations of PRC typically deployed semiconductor lasers~\cite{brunner2013parallel}, and optical amplifiers~\cite{duport2012all}, but now approaches based on photonic integrated circuit (PIC) components, such as microring resonators, have been demonstrated, highlighting the flexibility, versatility and scalability of the platform and PRC techniques~\cite{vandoorne2014experimental,donati2024time,donati2025alloptical}.

PRC architectures typically fall into one of two species, spatial (physically networked) or temporal (delayed-based)~\cite{abreu2024photonics}.
Temporal reservoirs leverage a single physical nonlinear node to create a network of many coupled virtual nodes via time-multiplexing~\cite{appeltant2011information}.
Coupling and memory in these systems is created through the short-term dynamics in the system, and the long-term feedback created through an optical delay line. 
Spatial systems, on the other hand, implement multiple discrete nonlinear nodes, that are physically coupled using various techniques (waveguides, diffractive optics, surface-interactions etc.), to create truly parallel reservoir nodes in 2D structures/arrays~\cite{rafayelyan2020large,antonik2019large,bueno2018reinforcement,pfluger2024experimental}.
Each architecture provides unique benefits; temporal reservoirs are more hardware-friendly, easily scalable, and serve well memory-intensive tasks; while spatial systems provide instantaneous input transformation for real-time operations.

The combination of spatial and temporal architectures has also been proposed in recent theoretical studies~\cite{rohm2018multiplexed,bauwens2024combining,huang2023Enhanced}.
These systems aim to compliment the high virtual node density of temporal reservoirs with the distributed complexity and inherent parallelism (speed) of spatial reservoirs.
Recent studies include modeling of a spatial PIC reservoir coupled to a delayed-based semiconductor laser reservoir~\cite{bauwens2024combining}; and a study on multiple temporal VCSEL reservoirs in parallel, with mutual coupling configurations~\cite{huang2023Enhanced}. On the experimental side, reservoirs employing both wavelength and temporal multiplexing for high throughput and computational efficiency has also been demonstrated~\cite{Lupo2023deepPRC,Aadhi2025scalable}.
However, beyond the time-multiplexing of matrix vector multiplication operations for deep neural networks~\cite{Chen2022Deep}, reports of experimental spatio-temporal reservoir architectures with an array of lasers is lacking. 
In this work, an experimental reservoir with both spatial and temporal multiplexing (spatio-temporal) is proposed based upon an array of diffractively-coupled VCSELs. 
The proposed spatio-temporal PRC system uses Iris flower classification as a benchmarking task, before the performance is analysed for increasing spatial and virtual (temporal) node counts.
This experimental work highlights that the performance of an already existing PRC can be enhanced using a hybrid operation approach, via the introduction of additional virtual nodes. 
This approach benefits from the fast dynamics of semiconductors lasers, imposing only a small trade-off in operation speed. 
The following Sections\,\ref{sec:level1} \& \ref{sec:level2}, introduce the experimental array-based PRC and present its classification performance for different configurations, respectively. 

\section{\label{sec:level1}EXPERIMENTAL SETUP: SPATIO-TEMPORAL VCSEL ARRAY PRC}

\begin{figure}[h!]
\includegraphics[width=0.3\textwidth]{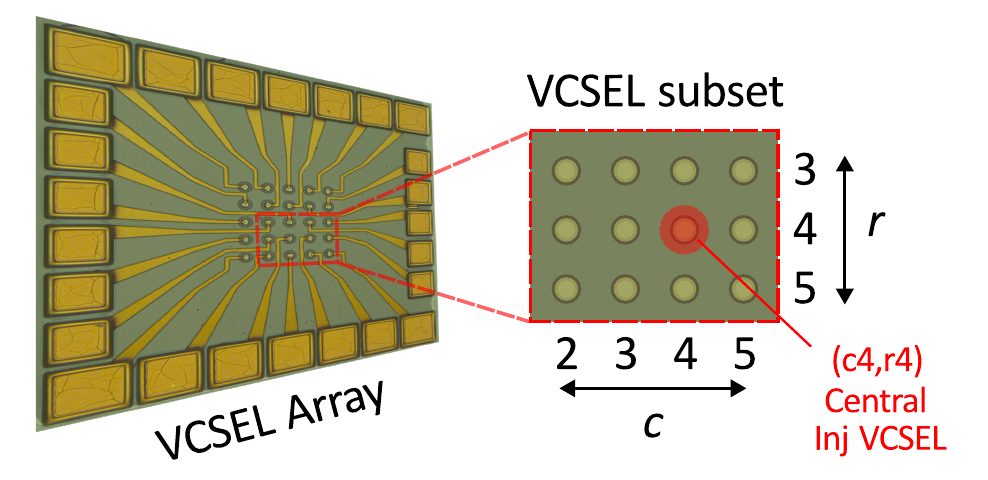}
\caption{\label{fig:VCSEL Naming} VCSEL array and nomenclature. Optical injection is centered on VCSEL ($c4,r4$). }
\end{figure}

In this work, the core non-linear photonic elements are VCSELs of a custom-manufactured VCSEL array, previously introduced in~\cite{Heuser2020,pfluger2024experimental,Pfluger2023Injectionlocking}. 
The VCSELs of the array are arranged in a $5 \times 5$ square lattice with a pitch of $p \approx 80\,{\mu m}$ between neighbouring VCSELs.
Each laser is electrically contacted allowing their pump current to be controlled independently.
The VCSELs incorporate AlGaAs/GaAs Distributed Bragg Reflector mirrors and an active region of GaInAs quantum wells with GaAsP barrier layers.
The aperture of the lasers are defined by an oxidised AlGaAs layer in the previously described mesa. The lasers emit with a dominant fundamental transverse mode at $\sim$\,976\,nm with a high degree of homogeneity in their polarisation orientation due to a slight elliptical cross-section. More fabrication details can be found in \cite{Heuser2020}.
For individual VCSELs, we define a nomenclature as shown in Fig.\,\ref{fig:VCSEL Naming}.
A VCSEL in column $X$ and row $Y$ of the array is referred to as VCSEL $(cX,rY)$.

\begin{figure}[h!]
\includegraphics[width=\linewidth]{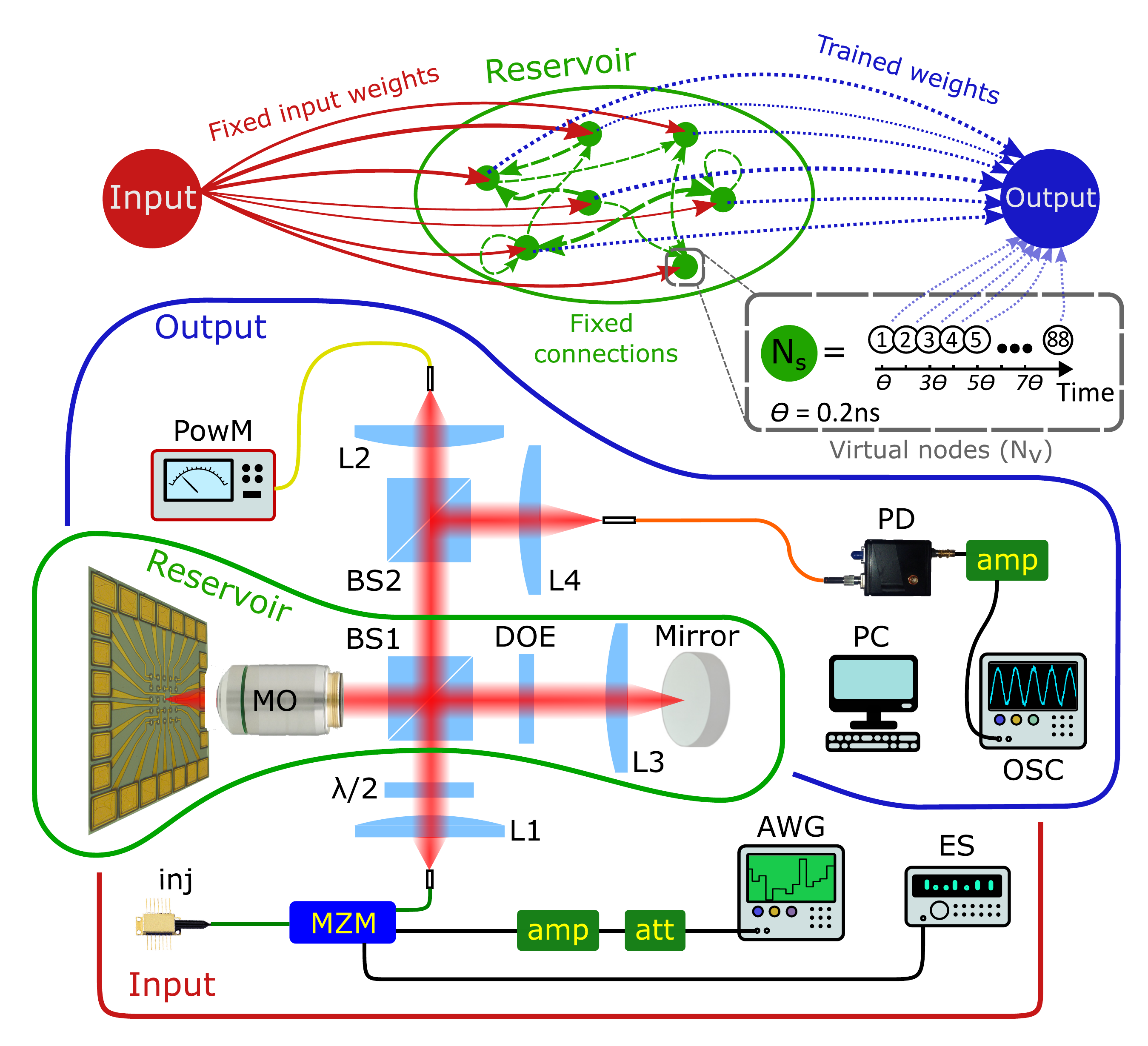}
\caption{\label{fig:setup} Schematic experimental setup. The injection branch consists of a DBR injection laser (inj), whose intensity is modulated using a Mach-Zehnder modulator (MZM), which in turn is biased using an electrical sourcemeter (ES). The output of an arbitrary waveform generator (AWG) passes an electrical attenuator (att) and an electrical amplifier (amp) before entering the MZM's radio-frequency (RF) port. The intensity-modulated light passes an aspheric lens (L1) and a half-wave plate ($\lambda/2$). Via reflection at a 50/50 beam splitter (BS1), the injected signal enters the external cavity and the VCSEL signal enters the analysis branch, which contains a 30T/70R beam splitter (BS2), an aspheric (L2) and a plano-convex lens (L4), a powermeter (PowM), a photodiode (PD), and an oscilloscope (OSC). The external cavity is formed by a microscope objective (MO), a lens (L3) and a mirror. A diffractive optical element (DOE) creates beam copies, establishing coupling between VCSELs and enabling simultaneous injection into all the VCSELs of the array. 
Green lines represent PM fiber, orange lines MM fiber, yellow lines SM fiber, and black lines coaxial cables. As illustrated in the to RC diagram, a spatio-temporal reservoir is created by masking the reservoir input and sampling each spatial node (VCSEL) for several virtual temporal nodes.}
\end{figure}

Figure\,\ref{fig:setup} shows the experimental setup of the array-based PRC and the proposed implementation of both spatial and virtual nodes. This setup uses light from an external injection laser (inj, Thorlabs DBR976PN), that passes a Mach-Zehnder modulator (MZM, EOSpace AZ-0K5-10-PFA-PFA-970), to input intensity-modulated information into the system.
The injection light is emitted from a polarisation-maintaining fiber tip and is collimated using an aspheric lens (L1), before it is polarisation-aligned to the VCSEL emission using a $\lambda /2$ waveplate.
An external feedback cavity, created using a microscope objective, a lens (L3) and a mirror is used to introduce recurrence and spatial coupling between VCSELs via a diffractive optical element (DOE). 
The injection enters the external cavity via a reflection at BS1 and passes the DOE twice before being injected into the VCSELs.
This results in the injection intensity being spatially split before simultaneously injecting multiple VCSELs (spatial nodes) at the same time.
An arbitrary waveform generator (AWG, Tektronix AWG7122B, 12\,GSa/s), whose signal first passes a tunable electrical attenuator (att) and then a broadband electrical amplifier (amp), controls the modulation of the MZM and the injection intensity.
The MZM is biased to the maximal positive slope of its $\sin^2$ nonlinearity by applying a voltage of $U_\mathrm{bias} = 7.9\,V$.

The VCSELs' emission enters the measurement arm after being reflected at BS1.
The output passing through the 30T/70R beam splitter (BS2) is coupled into a single-mode (SM) fiber using an aspheric lens (L2).
This SM fiber's position serves as a reference for optical alignment when maximising the fiber-coupled intensity measured with an optical powermeter (PowM).
The output reflected at BS2 is fiber-coupled into a multimode (MM) fiber using a plano-convex lens (L4).
Output light from the MM fiber is measured using a photodiode (PD, New Focus 1554-A-50PD, 10 kHz to 12 GHz bandwidth), before further amplification via another broadband electrical amplifier (amp). 
This amplification maximises the vertical resolution of the signal measured on a fast real-time oscilloscope (OSC, Lecroy Wavemaster 816Zi, 16\,GHz analog bandwidth, 40\,GS/s).

The VCSEL emission (spatial nodes) of the PRC are accessed individually by adjusting the position of the MM fiber.
For our study we selected a $4 \times $3 segment of the VCSEL array (from ($c2,r3$) to ($c5,r5$)) to keep the manual readout procedure and wavelength alignment requirements manageable.
The wavelength alignment of the VCSEL array is achieved by tuning the pump current of each VCSEL and by varying the temperature of the external injection laser.
The VCSELs of the array operated with multiple longitudinal modes. 
The highest wavelength mode of each VCSEL is aligned such that the mean wavelength of the array is 976.78\,nm.
VCSELs ($c3,r3$) and ($c4,r4$) produce the highest and lowest wavelengths at 976.851\,nm and 976.736\,nm, respectively, resulting in a 0.115\,nm spread of wavelength across the array. 
At 320.8\,K, with a pump current of 450\,mA, the inj laser wavelength is detuned $\sim$ -0.122\,nm (38.3\,GHz) from the mean wavelength of the array.
Previous studies have shown that, under these conditions, the VCSEL array elements can be stably injection-locked~\cite{Pfluger2023Injectionlocking}.
The free-space optical power of the modulated external laser entering the cavity is 7.3\,mW.  
Optical injection is centered on VCSEL ($c4,r4$), which due to its central position and direct reflection at BS2, makes its node output indistinguishable from the optical injection signal.
When the output of the central spatial node ($c4,r4$) is disregarded, the analysis of DOE-coupled reservoir is performed using a subset of network, i.e. the remaining 11 spatial nodes ($N_S$ = 11). 

\begin{figure}[h!]
\includegraphics[width=0.4\textwidth]{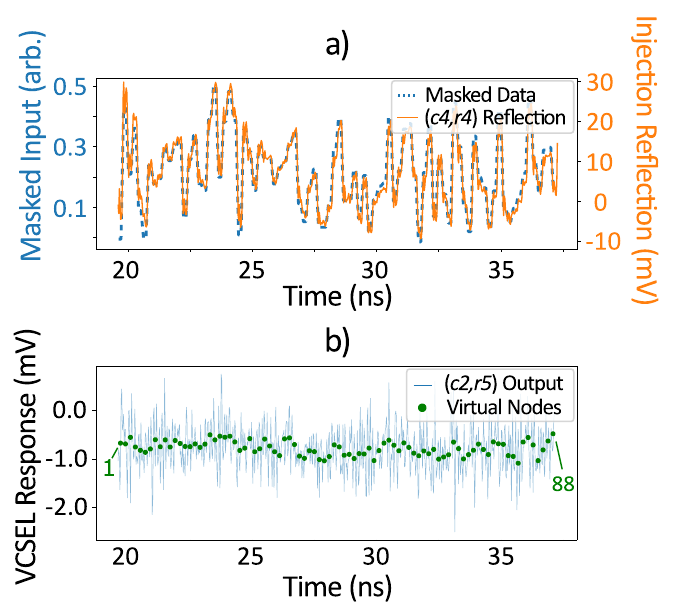}
\caption{\label{fig:Res In-Out} Reservoir Input and Output Timeseries. Top: 4 Features of 1 flower datapoint masked for 88 temporal nodes. The reflection from central injection VCSEL ($c4,r4$) shows the optical input of the reservoir. Bottom: The corresponding optical output of VCSEL ($c2,r5$) is shown with the extracted reservoir node value calculated by averaging 8 points. }
\end{figure}

To realise a spatio-temporal reservoir, the 11 detected spatial nodes are each driven by the time-multiplexed injection, defining 88 virtual temporal nodes ($N_v$ = 88).
These virtual nodes are created by masking every task feature with a random sequence of values ($0.2, 0.4, 0.6, 0.8$ and $1.0$), defining an 88 valued input vector for each input data sample.
Each of the input vector values are used to generate two samples at 10\,GSa/s in the AWG, corresponding to a virtual node separation of 200\,ps.
An example of a masked input datapoint is shown in Fig.\,\ref{fig:Res In-Out}(a), overlaid with the corresponding optical injection signal measured via the ($c4,r4$) spatial output with the VCSEL array off. 
The separation of the virtual nodes (200\,ps) was selected such that an integer number of virtual nodes (11) matched the external cavity feedback time of 2.2\,ns. The integer number provides synchronous node feedback that adds coupling between virtual nodes within each input data sample.  
Datapoints are time-multiplexed into the injection with a reset interval of 2.2\,ns after each datapoint, to clear the system memory and prevent cross contamination of data.
The 88 temporal nodes are determined from the acquired time traces of each VCSEL (with the OSC averaged over 250 measures).
Specifically, the virtual node readout is calculated as the average of 8 OSC samples to compensate for AWG/OSC sample rate mismatch ($ (10/2)/40 = 8$).
The measured output of spatial node ($c2,r5$) with its corresponding 88 extracted virtual nodes, is shown in Fig.\,\ref{fig:Res In-Out}(b).
Overall, the contribution from 11 different spatial nodes, with 88 virtual nodes in each, creates a spatio-temporal reservoir of 968 nodes. 

To test the classification performance of the spatio-temporal reservoir, Fisher’s iris flower classification dataset is used~\cite{FISHER1936}. 
Each datapoint within the dataset contains 4 features (the sepal and petal lengths and widths), with a total of 150 different specimens provided.
The dataset contains three species, or classes, of flower (Setosa, Virginica \& Versicolor). 
The dataset is balanced with 50 datapoints of each class, where only one of the three classes is linearly separable (Setosa - class 1).
The training of the spatio-temporal reservoir is performed offline with weight calculation via ridge regression (ridge coefficient $\alpha = 1\times 10^{-6}$) with 5 fold cross-validation (80\% for training, 20\% for testing).
The reservoir is trained with a 3 class label set for direct classification readout. 

\section{\label{sec:level2}Results}

The analysis of the system is performed by breaking down the influence of spatial and temporal node variation.
First, in Section\,\ref{sec:level2:solo}, the performance of each individual spatial node is considered.
Secondly, in Section\,\ref{sec:level2:Inc}, the performance of the system is considered when the number of spatial nodes is increased.
Finally, in Section\,\ref{sec:level2:Full}, the overall performance of the system is presented, showing full spatio-temporal operation of all 968 nodes. 
In each of these Sections, the number of virtual temporal nodes used for training is increased sequentially from a single node to all 88 nodes.   

\subsection{\label{sec:level2:solo}INDIVIDUAL SPATIAL NODE ANALYSIS}

\begin{figure}[h!]
\includegraphics[width=0.4\textwidth]{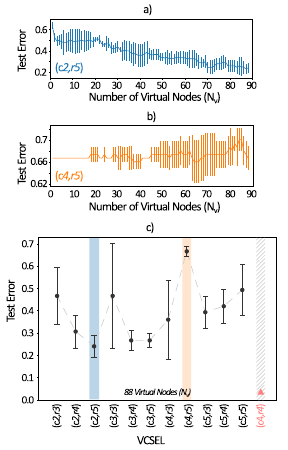}
\caption{\label{fig:solospatial}Performance of individual spatial nodes. Test error for increasing virtual nodes for both the highest (a) - ($c2,r5$) and lowest (b) - ($c4,r5$) performing spatial nodes. (c) Calculated test error of each VCSEL with 88 virtual nodes. Grey dashed highlight shows the disregarded central spatial node ($c4,r4$) (pink triangle).}
\end{figure}

First, following the injection of the temporally masked data (containing 88 virtual nodes in total), the performance of individual spatial nodes is analysed. 
In this case, the output virtual nodes from each VCSEL are considered as individual temporal reservoirs fed by a common optical input. 
Fig.\,\ref{fig:solospatial}(a) and (b) show the test classification error for VCSELs ($c2,r5$) and ($c4,r5$), when trained using an increasing number of their virtual (temporal) nodes.
The virtual nodes used to train the reservoir were selected in sequential order.
In Fig.\,\ref{fig:solospatial}(a), the test error slowly decreases with the increasing number of virtual nodes considered, reaching 0.24 $\pm$ 0.49 when using all 88 virtual nodes.
As seen in Fig.\,\ref{fig:solospatial}(c), this corresponds to the top performing spatial node in the system. 
In Fig.\,\ref{fig:solospatial}(b), the lowest performance node ($c4,r5$), shows a different trend, fluctuating around 0.66, due to the probability of guessing the class correctly (1 in 3). 
From Fig.\,\ref{fig:solospatial}(c), we can identify the spatial nodes that are operating in a parameter space more suitable for information processing.
This performance varies across the spatial nodes due to several experimental factors. 
The optical alignment of VCSEL wavelengths was prioritised over driving current, resulting in several nodes operating near lasing threshold, or with a large current, creating non-uniform non-linear transformations across the array. 
Similarly, due to the DOE inside the reservoir, different spatial nodes receive different levels of injection power, again influencing the non-linear transformation performed by the VCSELs.
Additionally, various driving currents result in different output powers, making signal reading and node extraction more difficult for nodes such as ($c4,r5$).
The individual performance of the central spatial node ($c4,r4$) is also shown (pink triangle) Fig.\,\ref{fig:solospatial}(c). 
The central node outperformed all other individual spatial nodes with an error of 0.033 $\pm$ 0.00 (using 88 virtual nodes), reaching errorless performance for low virtual node counts ($N_v$ = 11). 
We argue that the high optical power of the signal measured from the central node position is responsible for the unmatched performance of the central spatial node. 
This spatial node produces a measurement with a signal-to-noise ratio (SNR) superior to other spatial nodes, as clearly seen when comparing the output of the injection VCSEL ($c4,r4$) (Fig.\ref{fig:Res In-Out}(a)) to that of VCSEL ($c2,r5$) (Fig.\ref{fig:Res In-Out}(b)).
Additionally, optical input reflection at the central node position enables the use of the injection VCSEL signal for state augmentation, as the system is trained on both masked node inputs and reservoir-transformed node outputs. This feature, unique to the central spatial node, may contribute to the overall performance enhancement.
With the central spatial node disregarded, the individual performance analysis provides insight into the performance hierarchy of the remaining 11 spatial nodes in the current configuration. 

\subsection{\label{sec:level2:Inc}INCREASING SPATIAL NODE ANALYSIS}


Secondly, the performance of the system was analysed with an increasing number of spatial nodes ($N_s$).
Here, the 88 virtual nodes ($N_v$ = 88) from each VCSEL are pooled together to generate an increasingly large spatio-temporal reservoir.
Using the performances measured in Section \ref{sec:level2:solo}, two training sequences are created, one of decreasing spatial node performance (Order A, O-A), and one of increasing spatial node performance (Order B, O-B).
O-A begins training with the lowest error node ($c2,r5$), and O-B begins training with the highest error node ($c4,r5$).

\begin{figure}[h!]
\includegraphics[width=0.48\textwidth]{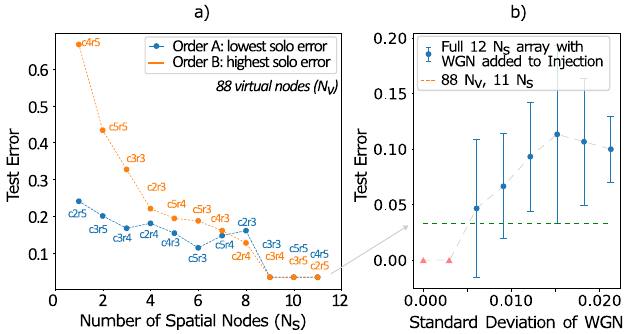}
\caption{\label{fig:Order} Performance with increasing number of spatial nodes, with 88 virtual nodes in each VCSEL. a) Spatial nodes are added in sequence of decreasing individual performance (O-A, blue) and increasing individual node performance (O-B, orange). b) Comparison of performance when all 12 spatial nodes (including the injection VCSEL) are considered (blue), to that of the 11 spatial node system (orange dashed). Increasing white gaussian noise (WGN) is added to the signal measured from injection VCSEL ($c4,r4$) to compensate for its large signal-to-noise ratio.}
\end{figure}

Fig.\,\ref{fig:Order} shows the dependence of spatio-temporal reservoir performance on the inclusion of additional spatial nodes (VCSELs).
Comparing both training sequences, Fig.\,\ref{fig:Order}(a) reveals that training initially with the best individual performing nodes will grant you higher performance for smaller networks.
Here, performance using $N_s$ = 3 with O-A, is almost equivalent to twice as many nodes ($N_s$ = 7) in O-B.
Interestingly, training in O-A does not guarantee the best performance for each $N_s$, as $N_s$ = 8 achieves higher performance using O-B.
Further, both Orders A \& B, become indistinguishable at $N_s$ = 9 with matching performance errors of 0.033 $\pm$ 0.042.
This indicates that in O-A the lowest performance nodes add no additional useful information.
However, O-A also shows that with additional nodes, the performance does not always improve, as you begin to add in nodes with less valuable or distracting information. 
In O-B, the system demonstrates that it can learn and classify information, just as well as O-A, without the top performing nodes, again showing a saturating performance for $N_s$ = 10.
Finally, at $N_s$ = 11, both O-A and O-B represent the same reservoir and performance. 
Fig.\,\ref{fig:Order}(a) reveals that reservoir performance does not improve uniformly with spatial node number ($N_s$) due to performance variations across the array.
Typically, training in order of highest individual performance will provide you with best performance for the smallest system.  

Additionally, the performance of the system when the signal measured from the central spatial node (injection VCSEL ($c4,r4$)) is included in the reservoir training is shown in Fig.\,\ref{fig:Order}(b). 
This reservoir, with a total of 1056 nodes ($N_v$ = 88, $N_s$ = 12), shows further performance enhancement with the increased spatial node count, reaching errorless classification (pink triangles).
However, to account for the large SNR of the injection VCSEL ($c4,r4$), we added white gaussian noise (WGN) of increasing standard deviation, and zero mean, to the measured signal before training. 
Fig.\,\ref{fig:Order}(b) shows that as the WGN noise increases, and the SNR of the injection VCSEL drops towards that of the other VCSELs in the array, the performance falls in line with that of the 11 spatial node reservoir (orange dashed).
Therefore, if the injection laser was measured with a similar optical power and noise level as the other VCSELs in the system, it would be unlikely to provide a significant performance boost. 
Similarly, higher SNR across the VCSEL array is expected to improve the overall performance of the spatiotemporal reservoir. 
While the inclusion of the central spatial node in the current configuration leads to a performance enhancement, this cannot be associated solely with an increasing number of spatial nodes in the reservoir, but more likely its larger SNR.


\subsection{\label{sec:level2:Full}FULL SPATIO-TEMPORAL ANALYSIS}

\begin{figure}[h!]
\includegraphics[width=\linewidth]{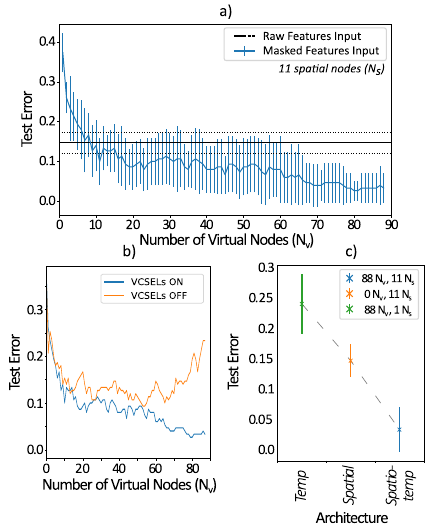}
\caption{\label{fig:TempPerformance} Full spatio-temporal reservoir performance using 11 spatial nodes. a) Influence of increasing the number of virtual nodes, compared to performance of unmasked feature injection (spatial-only reservoir). b) Spatio-temporal reservoir performance (with VCSELs on) compared to performance of measured masked inputs (reflections of injection with VCSEL off). c) Summary of results for different architecture performances.}
\end{figure}

Finally, the performance of the full spatio-temporal laser-array reservoir is shown in Fig.\,\ref{fig:TempPerformance} (a).
With the response of all 11 VCSELs ($N_s$ = 11) combined, the performance of the system is shown for increasing virtual node count ($N_v$). 
The test error of the system follows a downward trend with increasing $N_v$, reaching 0.033 $\pm$ 0.042 using all 88 virtual nodes from the 11 VCSELs. 
The highest performance of 0.026 $\pm$ 0.025 was achieved for 80 virtual nodes in each VCSEL.  
This performance can be compared to a spatial-only reservoir system by first looking at the system with a single virtual node.
The single virtual node spatio-temporal reservoir is equivalent to randomly masking your datapoint features together before injection into 11 VCSELs.
Performance with $N_v$ = 1, is shown in Fig.\,\ref{fig:TempPerformance} (a), where it reaches 0.37 $\pm$ 0.049.
Similarly, completing the task without virtual nodes (no temporal masking) is also achieved by sequentially injecting the unmasked datapoint features into all spatial nodes as floating point values, each held for a temporal node duration (0.2\,ns).
This input returns a benchmark performance of a spatial only reservoir (fed directly datapoint features), that we can also use to compare spatial and spatio-temporal performance.
In this case, the injection of unmasked features achieved a test error of 0.146 $\pm$ 0.027 (see black line in Fig.\,\ref{fig:TempPerformance} (a)).
The full spatio-temporal reservoir therefore outperforms both (masked and unmasked feature input) spatial-only reservoirs.

Fig.\,\ref{fig:TempPerformance} (b) shows the test error of the full spatio-temporal reservoir (blue line), alongside the task performance when trained on the measured masked input to all 11 VCSELs (orange line). 
For this result, a reflection of the injection at each spatial node position was measured with the VCSELs off.
Without VCSEL emission, training this optical input signal is equivalent to performing a linear regression on the masked data features.
Fig.\,\ref{fig:TempPerformance} (b) reveals that the VCSEL outputs are responsible for the improved performance, with the largest benefit occurring when using all virtual nodes.

Finally, Fig.\,\ref{fig:TempPerformance} (c) showcases the performance improvement when implementing a hybrid spatio-temporal reservoir compared to temporal and spatial-only architectures. The best performing spatial node ($c2,r5$) with 88 virtual nodes and the unmasked-feature trained (spatial-only) reservoir, are each outperformed by hybridising into a spatio-temporal laser-array architecture. Furthermore, despite the identification of several poor-performing VCSELs and the low SNRs achieved across the array, the performance improvement when combining spatial and temporal nodes is still significant.

\section{\label{sec:level3}Conclusion}
In conclusion, this work demonstrates a spatio-temporal VCSEL-array PRC and analyses the systems performance using the Iris flower classification task. 
The analysis reveals that by adding increasing numbers of virtual nodes to each spatial node, higher classification performance can be achieved.
Performance improvement is dependent on the effectiveness of spatial nodes, which was assessed by training the nodes individually, indicating which VCSELs were operating in a useful dynamical regime.
Further, it was shown that combining high-performing spatial nodes first, more quickly (using a smaller network) improved the overall spatio-temporal reservoir performance. 
Finally, the overall analysis reveals that the spatio-temporal reservoir could achieve performance with a minimum error of 0.026 $\pm$ 0.025 using a total of 880 nodes, or 0.033 $\pm$ 0.042 using all 968 nodes.
Importantly, the combined system outperformed both temporal and spatial-only versions of the system.
The implementation of this architecture in a photonic VCSEL array provides an easy pathway to push beyond reported multi-layered matrix-vector operations, and realise training-simplified reservoir recurrent neural networks. 
Further, given the high speed non-linearity in VCSELs, the functionality and performance gained is a favourable trade-off given only an additional 17.6\,ns (88 nodes at 0.2\,ns) is required for the hardware readout.

Looking beyond this experimental demonstration, several avenues exist to further optimise the spatio-temporal architecture. First, the current setup encounters limitations regarding the signal-to-noise ratio during the photodetection of individual VCSEL outputs. Future iterations incorporating integrated readout mechanisms would significantly enhance signal fidelity and system robustness. Second, the efficiency of the optical coupling is heavily dependent on precise wavelength matching between the injection laser and the array elements, a factor currently constrained by fabrication tolerances. Advances in device fabrication uniformity would improve this matching, ensuring that a larger proportion of spatial nodes operate within their optimal non-linear dynamical regimes without requiring extreme current tuning. Further, the polarisation of the VCSELs has so far been homogeneous  across the array (by design). In principle, the polarisation might be utilised to add another degree of freedom to each emitter. For individual VCSELs this has been demonstrated in \cite{Vatin2020DualTask}. Finally, the architectural balance between spatial parallelism ($N_s$) and temporal multiplexing ($N_v$) represents a flexible design parameter rather than a fixed constraint. The optimal compromise is ultimately application-dependent, offering the versatility to tailor future systems to prioritize either ultrafast processing speeds (favouring high spatial node counts) or maximized computational performance (leveraging deeper temporal reservoirs) to meet the specific demands of varying AI and computer vision tasks.

\begin{acknowledgments}
The authors would like to thank T. Heuser, J.A. Lott \& S. Reitzenstein for designing and fabricating the custom VCSEL array, and P. Canvelles and D. Brunner for the design of the control board. The authors acknowledge support from the UKRI Turing AI Acceleration Fellowships Programme (reference EP/V025198/1), EU Pathfinder Open project ‘SpikePro’ (Grant ID 101129904), UK Multidisciplinary Centre for Neuromorphic Computing (reference UKRI982), EPSRC Project ‘ProSensing’ (reference EP/Y030176/1), Spanish State Research Agency through the Severo Ochoa and Mar\'{\i}a de Maeztu Program for Centers and Units of Excellence in R\&D (No. CEX2021-001164-M) and the INFOLANET project (No. PID2022-139409NB-I00) funded by MCIN/AEI/10.13039/501100011033, and the EU’s Research and Innovation Programme under the Marie Skłodowska-Curie Grant Agreement No. 101169118 (POST DIGITAL+).
\end{acknowledgments}

\section*{Data Availability Statement}
For the purpose of open access, the author(s) has applied a Creative Commons Attribution (CC BY) licence to any Author Accepted Manuscript version arising from this submission. All data underpinning this publication are openly available from the University of Strathclyde KnowledgeBase at https://doi.org/10.15129/3598b6cd-6536-4a6f-8455-34cde308ad40.
\bibliography{Main}

\end{document}